\documentclass[11pt]{article}
\usepackage[a4paper,left=1.0in,right=1.0in,top=1.1in,bottom=1.3in]{geometry}

\usepackage{setspace}
\usepackage{microtype}

\usepackage{todonotes}
\usepackage{authblk}
\usepackage{xcolor}
\usepackage{etex,ifthen}
\usepackage{etoolbox}
\usepackage{graphicx}
\usepackage{rotating}
\usepackage{tikz}
\usepackage{pgfplots}\pgfplotsset{compat=1.14}
\usepackage{tikz-cd}
\usepackage{url}
\usepackage{calc}
\usepackage[font={small,it}]{caption}
\usepackage{complexity}
\usepackage{alphalph}
\usepackage{amsthm}
\usepackage{thmtools}
\usepackage{eqparbox}
\usepackage{afterpage}
\usepackage{amsfonts,amsmath,amssymb}
\usepackage{mathrsfs}
\usepackage{algorithm}
\usepackage{algpseudocode}
\usepackage{bm}
\usepackage[framemethod=TikZ]{mdframed}
\usepackage{listings}
\usepackage[english]{babel}
\usepackage{hyperref}
\usepackage{verbatim}
\usepackage{mathptmx}
\usepackage[all]{xy}

\usetikzlibrary{chains,fit}
\usetikzlibrary{shapes.geometric}
\usetikzlibrary{matrix,positioning,calc}
\usetikzlibrary{decorations.markings}
\usetikzlibrary{decorations.pathmorphing}
\tikzstyle{vertex}=[circle, draw, inner sep=0pt, minimum size=21pt]
\tikzstyle{svertex}=[circle, draw, inner sep=0pt, minimum size=3pt]
\tikzstyle{dvertex}=[circle, draw, inner sep=0pt, minimum size=9pt]
\tikzstyle{vertbox}=[draw, inner sep=0pt, minimum size=8pt]
\newcommand{\vertex}{\node[vertex]}

\newcommand{\oset}[3][0ex]{
    \mathrel{\mathop{#3}\limits^{
		\vbox to#1{\kern-2\ex@
		\hbox{$\scriptstyle#2$}\vss}}}}

\mdfdefinestyle{MyFrame}{%
    linecolor=black,
    innerbottommargin=\baselineskip,
    backgroundcolor=gray!10!white}

\newcommand{\Oh}{\mathscr{O}}
\newcommand{\iso}{\mathsf{iso}}

\newcommand{\cbef}{c_{\,\mathsf{before}}}
\newcommand{\caft}{c_{\,\mathsf{after}}}

\declaretheorem[numberlike=equation]{Theorem}
\declaretheorem[numberlike=equation]{Lemma}

\declaretheorem[numberlike=equation]{Proposition}

\declaretheoremstyle[bodyfont=\it,qed=$\lozenge$]{defstyle}
\declaretheorem[numberlike=equation,style=defstyle]{Definition}
\declaretheorem[numberlike=equation,style=defstyle]{Remark}
\declaretheorem[numberlike=equation,style=defstyle]{Example}
\declaretheorem[numberlike=equation]{Claim}
\declaretheorem[numberlike=equation]{Fact}

\makeatletter
\patchcmd{\ALG@step}{\addtocounter{ALG@line}{1}}{\refstepcounter{ALG@line}}{}{}
\newcommand{\ALG@lineautorefname}{Line}
\makeatother

\addto\extrasenglish{%
}

\hypersetup{
	colorlinks,
	linkcolor=red!70!black,
	citecolor=blue!70!black,
	urlcolor=green!70!black
}

\title{The Space Complexity of Sum Labelling}

\author[1]{Henning Fernau\thanks{\texttt{fernau@uni-trier.de}, ORCiD: [0000-0002-4444-3220]}}
\author[2]{Kshitij Gajjar\thanks{\texttt{kshitij@comp.nus.edu.sg}, ORCiD: [0000-0003-0890-199X]}}

\affil[1]{Universit\"at Trier, FB 4 -- Informatikwissenschaften, Trier, Germany}
\affil[2]{National University of Singapore, 21 Lower Kent Ridge Rd, Singapore}

\begin{document}
	
	
	\maketitle
	
	
	\begin{abstract}
	A graph is called a~\emph{sum graph} if its vertices can be labelled by distinct positive integers such that there is an edge between two vertices if and only if the sum of their labels is the label of another vertex of the graph. Most papers on sum graphs consider combinatorial questions like the minimum number of isolated vertices that need to be added to a given graph to make it a sum graph. In this paper, we initiate the study of sum graphs from the viewpoint of computational complexity. Notice that every $n$-vertex sum graph can be represented by a sorted list of $n$ positive integers where edge queries can be answered in $\Oh(\log n)$ time. Therefore, limiting the size of the vertex labels upper-bounds the space complexity of storing the graph in the database.
	
    We show that every $n$-vertex, $m$-edge, $d$-degenerate graph can be made a sum graph by adding at most $m$ isolated vertices to it, such that the size of each vertex label is at most $\Oh(n^2d)$. This enables us to store the graph using $\Oh(m\log n)$ bits of memory. For sparse graphs (graphs with $\Oh(n)$ edges), this matches the trivial lower bound of $\Omega(n\log n)$. As planar graphs and forests have constant degeneracy, our result implies an upper bound of $\Oh(n^2)$ on their label size. The previously best known upper bound on the label size of general graphs with the minimum number of isolated vertices was $\Oh(4^n)$, due to Kratochv\'il, Miller \& Nguyen (2001). Furthermore, their proof was existential, whereas our labelling can be constructed in polynomial time.
	\end{abstract}
	
	\newpage


\section{Introduction}

There is a vast body of literature on graph labelling, testified by an ever-expanding survey on the topic maintained by Gallian~\cite{Gal2020}. 
The 553-page survey (as of December 2020) mentions over 3000 papers on different ways of labelling graphs. In this paper, we focus on a type of labelling introduced by Harary~\cite{harary} in 1990, called sum labelling.

\begin{figure}
\begin{centering}
\begin{tikzpicture}

\def \xoff {6};
\def \isoso {1};

\node[vertex](a1) at (-4, 4) {};
\node[vertex](a2) at (-3, 3) {};
\node[vertex](a3) at (-3, 5) {};

\node[vertex](b1) at (-4+\xoff,4) {$1$};
\node[vertex](b2) at (-3+\xoff,3) {$3$};
\node[vertex](b3) at (-3+\xoff,5) {$2$};

\node[vertex](iso1b) at (-3+\xoff+\isoso,3.4) {$5$};
\node[vertex](iso2b) at (-3+\xoff+\isoso,4.6) {$4$};

\node[vertex](d1) at (-4+2*\xoff,4) {$1$};
\node[vertex](d2) at (-3+2*\xoff,3) {$4$};
\node[vertex](d3) at (-3+2*\xoff,5) {$3$};

\node[vertex](iso1d) at (-3+2*\xoff+\isoso,3.4) {$7$};
\node[vertex](iso2d) at (-3+2*\xoff+\isoso,4.6) {$5$};

\node[] at (-3,2) {\Large (a)};
\node[] at (-3+\xoff,2) {\Large (b)};
\node[] at (-3+2*\xoff,2) {\Large (c)};

\begin{scope}[every path/.style={-}, every node/.style={inner sep=1pt}]
       \draw (a1) -- (a2) -- (a3) -- (a1);
       \draw (b1) -- (b2) -- (b3) -- (b1);
       \draw (d1) -- (d2) -- (d3) -- (d1);
       
\end{scope} 
\end{tikzpicture}
\caption{(a) This graph is not a sum graph, as it has no isolated vertices (\autoref{fact:sumgraphiso}); (b) This is an incorrect sum labelling of a sum graph, because the vertices labelled $1$ and $4$ are not adjacent but there is a vertex labelled $1+4=5$ in the graph (\autoref{def:sumlab}); (c) This is a correct sum labelling of a sum graph.}
\label{fig:validinvalidsumlabelling}
\end{centering}
\end{figure}
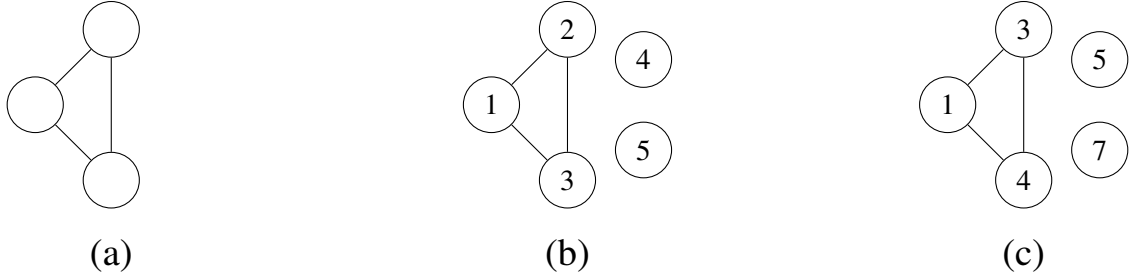

\begin{Definition} \label{def:sumlab}
A simple, undirected, unweighted graph $G$ is called a sum graph if there exists an injective function $\lambda: V(G)\rightarrow\mathbb{N}$ such that for all vertices $v_1,v_2\in V(G)$, $$(v_1,v_2)\in E(G)\quad \Longleftrightarrow\quad\exists\,v_3\in V(G)\ \text{ s.t. }\ \lambda(v_1)+\lambda(v_2)=\lambda(v_3).$$ Then we say that $\lambda$ is a sum labelling of (the vertices of) $G$.
\end{Definition}
Notice that~\autoref{def:sumlab} implies that given only the function $\lambda$ on the vertex set of a sum graph $G$, the edge set of $G$ can be obtained. Thus, $\lambda$ encodes the graph $G$. \autoref{fig:validinvalidsumlabelling} illustrates a helpful example to better understand sum labellings. The following elementary fact about sum graphs is fundamental to almost all research done so far on sum graphs, including ours.

\begin{Fact} \label{fact:sumgraphiso}
Every sum graph has at least one isolated vertex (a vertex of degree zero).
\end{Fact}

\begin{proof}
We will prove this fact by contradiction. Suppose there exists a sum graph $G$ without an isolated vertex. Let $L$ be the maximum label of a vertex in a sum labelling of $G$. As $L$ is not the label of an isolated vertex, there is a vertex adjacent to it. Let the label of the adjacent vertex be $x$. Then there exists a vertex with label $L+x$ in $G$ (\autoref{def:sumlab}), contradicting the fact that $L$ is the maximum label.
\end{proof}

Gould \& R\"odl~\cite{GouRod8891} showed that every $n$-vertex graph can be made a sum graph by adding at most~$n^2$ isolated vertices to it. In fact, certain graphs can be encoded much more succinctly with sum labelling than with the more traditional methods of storing a graph (e.g., adjacency matrix, incidence matrix, adjacency list). 
This makes sum labelling an intriguing concept not just to mathematicians but also to computer scientists. Sum labelling could also be of interest in graph databases~\cite{angles2008survey,angles2012comparison,kumar2015graph} and in collections of benchmark graphs~\cite{lancichinetti2008benchmark,dominguez2010survey,jouili2013empirical}. However, no systematic study of this question has been undertaken so far. With this paper, we intend to start such a line of research, bringing sum labellings closer to the research in~\emph{labelling schemes}~\cite{KanNaoRud92}. To the best of our knowledge, the only known application of sum labelling before our work is in secret sharing schemes~\cite{SlaSugMil2006}.

The idea of using sum labelling to efficiently store graphs was already considered by Sutton~\cite{Sut2000}. However, Sutton focused on the number of additional isolated vertices needed to store a given graph, whereas our focus is on the number of~\emph{bits} needed to store the graph.

In other words, while Sutton's work attempts to minimize the number of additional vertices, it does not take into account the~\emph{size} of the vertex labels required to do so. This is crucial because it is known that there are several graph families for which the size of the vertex labels grows exponentially with the number of vertices. One popular example is the sum labelling scheme for trees presented by Ellingham~\cite{ellingham}. Another example is the more esoteric graph family known as the generalised friendship graph~\cite{FerRyaSug2008}.

Another parameter associated with sum graphs is the difference between the largest and smallest label, called~\emph{spum} (also called~\emph{range} in~\cite{KraMilNgu2001}). Interestingly, while the concept of spum was around for quite some time (Gallian's survey~\cite{Gal2020} refers to an unpublished manuscript by a group of six students), the first publication that studies spum for various basic classes of graphs is a very recent one~\cite{SinTiwTri2021}. Unfortunately, this measure also does not reflect the whole truth about storing graphs, as it neglects the number of additional vertices that need to be stored. Moreover, spum is somewhat dependent on the definition of the sum number (see below for a formal definition), which might be slightly unnatural for the purpose of storing a graph.

In this paper, we also introduce a new graph parameter $\sigma_\textbf{store}$ that takes into account both the number of additional vertices and their label size. We explain this formally in the next section.

\section{Definitions and Main Result}

Let us now fix some notation in order to formally introduce the concepts in this paper.%
All our graphs are undirected, unweighted and simple, specified as $G=(V,E)$, where $V$ is the set of vertices and $E$ is the set of edges. If a vertex $v$ is an endpoint of an edge $e$, then we say that $v$ and $e$ are incident. The number of edges incident to a vertex is called its degree. 

\subsection{Sums and Spums}

As isolated vertices (or simply, isolates) are usually irrelevant in most practical applications, $\lambda$ (where $\lambda$ is a sum labelling of a sum graph $G$) can be also viewed as a description of $G\setminus I$, where $I$ is the set of isolates of~$G$. Then, $\lambda$ is called the~\emph{sum number encoding} of $G\setminus I$. Conversely, given a graph $G$ without isolates, the minimum number of isolates needed to be added in order to turn $G$ into a sum graph is called the~\emph{sum number} of $G$, written $\sigma(G)$, i.e., $G+\overline{K_{\sigma(G)}}$ is a sum graph. Here, $+$ denotes the disjoint union of graphs,
$\overline{H}$ denotes the complement of graph $H$, 
and $K_n$ is the complete graph on $n$ vertices. Thus, $\overline{K_n}$ is the empty (edgeless) graph on $n$ vertices. The~\emph{spum} of~$G$, written $\text{spum}(G)$, is defined as the minimum over all sum labellings of $G+\overline{K_{\sigma(G)}}$ of the difference between the 
maximum and minimum labels.

\subsection{The Size of Sum Number Encodings of Graphs}

A labelling function $\lambda$ can be also seen as operating on edges by the summability condition. $\lambda(e)$ for an edge $e=xy\in E$ is defined as $\lambda(x)+\lambda(y)$. A labelling of a sum graph $G=(V,E)$ is called an~\emph{exclusive} sum labelling~\cite{MilPRSST2005,MilRyaRyj2017,Rya2009,TugMil2003} if for every $e\in E$, we have $\lambda(e)=\lambda(i)$ for some isolate $i\in I\subseteq G$. Accordingly, $\epsilon(G)$ denotes the~\emph{exclusive sum number} of $G$, which is the minimum number of isolates to be added to $G$ such that $G+\overline{K_{\epsilon(G)}}$ is a sum graph that allows an exclusive sum labelling. Clearly, $\sigma(G)\leq\epsilon(G)$.

Are substantial savings possible when considering sum number encodings of graphs? As most research in the area of sum labellings went into studying quite specific families of graphs, some partial answers are possible. For instance, analyzing the expositions 
in~\cite{Pya2001,WanLiu2001}, one sees that for the complete bipartite graph $K_{n,n+1}$, with $n$ vertices in one partition and $n+1$ vertices in the other, $\sigma(K_{n,n+1})=2n-1$. In other words, we need $4n$ numbers in order to represent $K_{n,n+1}$. Ignoring the size of these numbers,
this is a clear advantage over any traditional way to store $K_{n,n+1}$, which would need $\Oh(n^2)$ bits when using adjacency matrices and even $\Oh(n^2\log n)$ bits when using adjacency lists. However, after a closer look at the labelling presented in~\cite{Pya2001}, it becomes clear that the numbers needed to label a $K_{m,n}$ are of size $\Oh(nm)$. Therefore, storing the complete bipartite graph $K_{n,n+1}$ needs only $\Oh(n\log n)$ bits, using its sum graph encoding. As we will see later, this is in fact storage-optimal, in a certain sense.

Similarly, $\sigma(K_n)=2n-3$ is known for $n\geq 4$, i.e., $3n-3$ numbers are necessary to store the information about the complete graph $K_n$, while again traditional methods would need $\Oh(n^2)$ bits at least. As mentioned in~\cite{Smy91}, this can be obtained by labelling vertex $x_i$ with $4i-3$, with $1\leq i\leq n$, leading to isolate labels $4j+2$ for $1\leq j\leq  2n-3$. Hence, the sizes of the labels are in fact linear in~$n$, which is, in a sense, even better than what is known for complete bipartite graphs. We will continue our discussions on storage issues in the next section.
%
%
%
%
It is known that the sum number of  general graphs grows with the order of its edges~\cite{NagMilSla2001}. In fact, this can happen even with sparse graphs~\cite{HarSmy95,SutMil01}.

As we have seen so far, neither the sum number of a graph nor the spum of a graph models the storage requirements of storing graphs with the help of sum numberings in a faithful manner. Therefore, we suggest another graph parameter, based on 
\begin{equation}\label{eq-storage}
\textbf{storage}(\lambda,G)=\sum_{v\in V}\left\lceil\log_2(\lambda(v))\right\rceil\leq |V|\cdot\max_{v\in V}\left\lceil\log_2(\lambda(v))\right\rceil
\end{equation}
for a labelling $\lambda:V\to\mathbb{N}$ of a sum graph $G=(V,E)$. (Notice that one can store variable-size numbers using at most twice as many bits when compared to~\autoref{eq-storage} with Elias prefix codes~\cite{Eli75}.) Now, define $$\textbf{storage}(G)=\min\{\textbf{storage}(\lambda,G)\mid \exists \lambda:V\to\mathbb{N}: \lambda \text{ is a sum labelling of }G\}\,.$$
Then, for an arbitrary graph $G'=(V',E')$ one could define $$\sigma_\textbf{store}(G')=\min\{\textbf{storage}(G)\mid \exists s\in\mathbb{N}: G=G'+\overline{K_s} \text{ is a sum graph}\}\,.$$ For instance, Ellingham's proof can be used to state: for an $n$-vertex tree $T$, Ellingham's construction leads to $\sigma_\textbf{store}(T)\in\Oh(n^2)$. This should be compared to any standard representation of trees that obviously needs $\Oh(n\log (n))$ space. However, our results prove that also with sum label representations, this upper bound can be obtained. In our construction, it is crucial that we also consider labellings that do not necessarily lead to a minimum sum number. This is also a difference concerning the definition of spum. As we are mostly interested in upper-bounding $\sigma_\textbf{store}(G')$ in this paper, we mainly discuss
$$\sigma_\textbf{store}^\text{max}(G')=\min\{\textbf{storage}^\text{max}(G)\mid \exists s\in\mathbb{N}: G=G'+\overline{K_s} \text{ is a sum graph}\}\,,$$
where for a sum graph $G=(V,E)$,
$$\textbf{storage}^\text{max}(G)=\min\{\textbf{storage}^\text{max}(\lambda,G)\mid \exists \lambda:V\to\mathbb{N}: \lambda \text{ labels }G\}\,,$$ with
$$\textbf{storage}^\text{max}(\lambda,G)= |V|\cdot \max_{v\in V}\left\lceil\log_2(\lambda(v))\right\rceil=|V|\cdot\left\lceil\log_2\left(\max\lambda(V)\right)\right\rceil\,.$$ By~\autoref{eq-storage}, $\sigma_\textbf{store}(G')\leq \sigma_\textbf{store}^\text{max}(G')$. 
A reader who likes to get more familiar with these notions is invited to first go through the next section. 

\smallskip
However, let us first 
state the main result of this paper, as we are now ready for it.


\begin{mdframed}[style=MyFrame,backgroundcolor=blue!1!white]
    \vspace{-0.5em}
	\begin{Theorem}[Main Result]\label{thm-main} Let $G'$ be a graph on $n$ vertices and $m$ edges with minimum degree at least one. Then, $\sigma_\mathbf{store}^{\max}(G')\in \Oh(m\cdot \log(n))$. More specifically, $$\sigma_\mathbf{store}^{\max}(G')\leq 9m(\log_2(n)+1)$$ for general graphs and  $$\sigma_\mathbf{store}^{\max}(G')\leq 3m(2\log_2(n)+\log_2(12d))<3dn(2\log_2(n)+\log_2(12d))$$ for $d$-degenerate graphs. Furthermore, the sum labelling can be computed in polynomial time.
    \end{Theorem}
	\vspace{-0.5em}
\end{mdframed}

In particular, this means that $\Oh(n\log(n))$ bits are sufficient to store trees with sum labellings, as they are 1-degenerate graphs. A similar result holds for planar graphs, as they are 5-degenerate. We show that these bounds are optimal for storing graphs, up to constant factors. We also relate to the literature on adjacency labelling schemes (see, e.g.,~\cite{KanNaoRud92,Pel2000}, or more recently,~\cite{BonGavPhi2020,DujEGJMM2020}).

To give a flavour of our algorithm, notice that it also works in the streaming or online setting, in which vertices are being given one-by-one by an adversary. 

\section{Sum Labelling a Disjoint Collection of Edges}

This section should be treated as an introductory exercise on sum labelling, and has no bearing on our main result. A reader familiar with sum labelling schemes may skip to the next section, possibly apart from the very last lines of this section.

It is known that trees have sum number~1; according to a remark following Theorem 5.1 in~\cite{ellingham}, this result translates to forests. However, the label sizes may grow exponentially in these constructions. As a warm-up and to explain the difficulties encountered while designing sum labellings, we present some constructions that label a disjoint collection of edges, or more mathematically speaking, a 1-regular graph, which we denote by $M_n$ (a matching on $n$ vertices, where $n$ is an even number).

\subsection{Exponential Solution (\autoref{fig:disjointedges} (a))}
If we have $n$ vertices (hence $n/2$ edges), we label the first edge $(2,3)$, the second one starts with the sum of the labels of the previous edge followed by its successor, i.e., $(5,6)$. Then we add up the previous two labels, continue with the successor, and so on. This can be brought into the following sum labelling scheme for $1$-regular graphs.
\begin{equation}\lambda(n)=\left\{\begin{array}{ll}
     2&  \text{if }n=1\\
     \lambda(n-1)+1& \text{if $n$ is even}\\
     \lambda(n-2)+\lambda(n-1)& \text{if $n$ is odd and }n>1
\end{array}\right.
    \label{eq:exponentialapproachtoMn}
\end{equation}

\begin{Lemma}
\label{lem:exponentialapproachtoMn}
For the labelling defined in~\autoref{eq:exponentialapproachtoMn}, we have $\lambda(n)\in\Theta\left(\sqrt{2}^{\;n}\right)$.
\end{Lemma}

\begin{proof}
The Online Encyclopedia of Integer Sequences suggests that this is another variation on Ulam numbers~\cite{OEIS-A002858} if we think of the starting point to be $\lambda(0)=1$. Then, 
$\lambda(n)$ (for $n>1$) can be seen as the smallest (when $n$ is even) or largest (when  $n$ is odd)  number larger than  $\lambda(n-1)$ that is a unique sum of two distinct earlier terms of the sequence. This connection suggests the following closed form:
$$\lambda(n)=\left\{\begin{array}{ll}
     3\cdot 2^{k-1}& \text{if $n$ is even, i.e., }n=2k\\
     3\cdot 2^k-1& \text{if $n$ is odd, i.e., } n=2k+1
\end{array}\right.$$
In other words, we have $\lambda(n)\in\Theta\left(\sqrt{2}^{\;n}\right)$, implying that $\lambda$ increases exponentially with $n$.
\end{proof}
Using this lemma, we can also conclude that $\textbf{storage}(\lambda,M_n)\in\Theta(n^2)$ for this labelling $\lambda$.

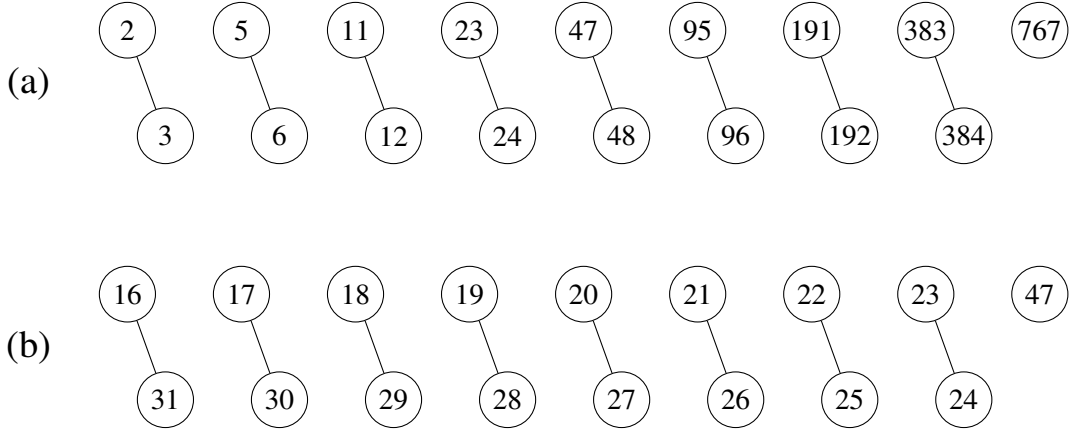
\begin{figure}
\begin{centering}
\begin{tikzpicture}

\def \yoff {3.5};

\node[vertex](a2) at (-3.5,1.4) {$16$};
\node[vertex](a1) at (-3,0) {$31$};

\node[vertex](b1) at (-2,1.4) {$17$};
\node[vertex](b2) at (-1.5,0) {$30$};

\node[vertex](c1) at (-0.5,1.4) {$18$};
\node[vertex](c2) at (0,0) {$29$};

\node[vertex](d1) at (1,1.4) {$19$};
\node[vertex](d2) at (1.5,0) {$28$};

\node[vertex](e1) at (2.5,1.4) {$20$};
\node[vertex](e2) at (3,0) {$27$};

\node[vertex](f1) at (4,1.4) {$21$};
\node[vertex](f2) at (4.5,0) {$26$};

\node[vertex](g1) at (5.5,1.4) {$22$};
\node[vertex](g2) at (6,0) {$25$};

\node[vertex](h1) at (7,1.4) {$23$};
\node[vertex](h2) at (7.5,0) {$24$};

\node[vertex](iso1) at (8.5,1.4) {$47$};

\node[] at (-4.8,0.7) {\Large (b)};
\node[] at (-4.8,0.7+\yoff) {\Large (a)};

\node[vertex](aa2) at (-3.5,1.4+\yoff) {$2$};
\node[vertex](aa1) at (-3,0+\yoff) {$3$};

\node[vertex](bb1) at (-2,1.4+\yoff) {$5$};
\node[vertex](bb2) at (-1.5,0+\yoff) {$6$};

\node[vertex](cc1) at (-0.5,1.4+\yoff) {$11$};
\node[vertex](cc2) at (0,0+\yoff) {$12$};

\node[vertex](dd1) at (1,1.4+\yoff) {$23$};
\node[vertex](dd2) at (1.5,0+\yoff) {$24$};

\node[vertex](ee1) at (2.5,1.4+\yoff) {$47$};
\node[vertex](ee2) at (3,0+\yoff) {$48$};

\node[vertex](ff1) at (4,1.4+\yoff) {$95$};
\node[vertex](ff2) at (4.5,0+\yoff) {$96$};

\node[vertex](gg1) at (5.5,1.4+\yoff) {$191$};
\node[vertex](gg2) at (6,0+\yoff) {$192$};

\node[vertex](hh1) at (7,1.4+\yoff) {$383$};
\node[vertex](hh2) at (7.5,0+\yoff) {$384$};

\node[vertex](isoiso1) at (8.5,1.4+\yoff) {$767$};

\begin{scope}[every path/.style={-}, every node/.style={inner sep=1pt}]
       \draw (a1) -- (a2); \draw (aa1) -- (aa2);
       \draw (b1) -- (b2); \draw (bb1) -- (bb2);
       \draw (c1) -- (c2); \draw (cc1) -- (cc2);
       \draw (d1) -- (d2); \draw (dd1) -- (dd2);
       \draw (e1) -- (e2); \draw (ee1) -- (ee2);
       \draw (f1) -- (f2); \draw (ff1) -- (ff2);
       \draw (g1) -- (g2); \draw (gg1) -- (gg2);
       \draw (h1) -- (h2); \draw (hh1) -- (hh2);
       
\end{scope} 
\end{tikzpicture}
\caption{(a) Labelling $M_{16}$ using~\autoref{eq:exponentialapproachtoMn}; (b) Labelling $M_{16}$ using~\autoref{eq:linearapproachtoMn}.}
\label{fig:disjointedges}
\end{centering}
\end{figure}

\subsection{Linear Solution (\autoref{fig:disjointedges} (b))}
Consider the following sum labelling scheme for 1-regular graphs $M_n$ on $n$ vertices. (We group endpoint labels of each edge together by parentheses.)
\begin{equation} \label{eq:linearapproachtoMn}
    (n,2n-1),(n+1,2n-2),\ldots,\left(\frac{3n}{2}-1,\frac{3n}{2}\right).
\end{equation}
All edge labels sum up to $3n-1$, which is the label of the isolated vertex. Also, it easy to see that these edges are the only ways in which two of the given $n$ numbers can sum to $3n-1$. Finally, even the sum of the two smallest labels between non-adjacent vertices (i.e., $n+(n+1)=2n+1$) is larger than the label of any other non-isolated vertex in the graph, proving that this is a valid sum labelling. As each label is in $\Theta(n)$, the overall space requirement of this labelling scheme is $\Theta(n\log(n))$. Moreover, as we can also see with the first labelling scheme,
$\sigma(M_n)=1$. Also, in contrast to the first scheme, this labelling scheme is exclusive. Hence, this approach also shows that $\epsilon(M_n)=1$. Finally, as the labels only grow linearly with $n$ with this labelling $\lambda$, we can also conclude that $\textbf{storage}(\lambda,M_n)\in\Theta(n\log(n))$.

\subsection{Disjoint Union of Several Identical Components} The previous consideration was quite special to 1-regular graphs. We now develop an argument that can be generalised towards a certain type of graph operation.
One can think of $M_n$ as being the disjoint graph union of $n/2$ times $M_2$.
For simplicity of the exposition, assume $n/2=2^d$ in the following. Label the vertices
$(v_{1,1},v_{2,1}),(v_{1,2},v_{2,2}),\dots,(v_{1,2^d},v_{2,2^d})$ of $M_n$ as follows, for $j=1,\dots,2^d$:
\begin{align*}
    \lambda(v_{1,j})&=1+8\cdot (j-1)+         2^{4+d}\cdot (2^{d}-j)\\
    \lambda(v_{2,j})&=2+8\cdot (2^{d}-j) +2^{5+d}\cdot (j-1)
\end{align*}
For instance, for $d=2$, we get $\lambda(v_{1,1})=1+8\cdot 0+64\cdot 3$,
$\lambda(v_{2,1})=2+8\cdot 3+64\cdot 0$, so that the connecting edge is testified
by the isolate label $3+8\cdot 3+64\cdot 3=219=(11011011)_2$. Also, $\lambda(v_{1,2})=1+8\cdot 1+64\cdot 2$, $\lambda(v_{2,2})=2+8\cdot 2+64\cdot 1$,
adding up again to 219. Likewise, $\lambda(v_{1,3})=1+8\cdot 2+64\cdot 1$,
$\lambda(v_{2,3})=2+8\cdot 1+64\cdot 2$, and finally $\lambda(v_{1,4})=1+8\cdot 3+64\cdot 0$ and $\lambda(v_{2,4})=2+8\cdot 0+64\cdot 3$. By construction, all numbers need at most $2d+4$ bits for labelling $2^{d+1}$ vertices. Hence, the overall space requirement for storing $M_n$ is again $\Oh(n\log(n))$ bits.

The zero bit introduced in the third and sixth binary position in the example ensures that the labels of two non-adjacent vertices cannot add up to the label of another vertex. This technique can be easily generalised to obtain the following result.

\begin{Lemma}\label{lem:disjointiterativeunion}
Let $G$ be a graph. Then, the $n$-fold disjoint graph union $G_n$ of $G$ with itself obeys  $\sigma_\textbf{store}(G_n)\in\Oh(n\log(n))$. Moreover, $\sigma(G_n)\leq\sigma(G)$.
\end{Lemma}

\section{Storing Graphs using Sum Labelling}



\paragraph{Alternative Notions.}
One of our motivations to return to sum labellings was the idea that one can use them to store graphs space-efficiently. This idea was already expressed in~\cite{KraMilNgu2001}. There, they consider the notion of the~\emph{range} of a sum graph $G$ that is realizing $\sigma(G')$, which happens to coincide with the notion called spum later.
But following this motivation (to store graphs), let us define the~\emph{range} of a
labelling $\lambda$ of a sum graph $G=(V,E)$ as the difference between $\max \lambda(V)$ and $\min\lambda(V)$. The idea behind is that it would suffice to store the numbers $\lambda(v)-\min\lambda(V)$ for all vertices $v\in V$, plus the value of 
$\min\lambda(V)$ once, instead of storing all values $\lambda(v)$, which could help us save some bits.

The following lemma tells us that this variation in our considerations (which could also lead to variations of the our definition of $\sigma_\textbf{store}$ and related notions) is not essential for our current considerations, as we mostly neglect constant factors. In particular, we might consider $|V|\cdot \left\lceil\log_2(\max \lambda(V)-\min\lambda(V))\right\rceil+\left\lceil\log_2(\min\lambda(V))\right\rceil$ as a more appropriate definition of the maximum estimate of  the storage requirements of a sum graph $G=(V,E)$ with respect to a sum labelling~$\lambda$.

\begin{Lemma}\label{lem:spum}
Let $\lambda$ be a sum labelling of a non-empty sum graph $G=(V,E)$, and let $\operatorname{range}(\lambda(V))=\max\lambda(V)-\min\lambda(V)$.
Then,
\begin{align*}
\operatorname{range}(\lambda(V))&>\min\lambda(V);\\ 2\cdot\operatorname{range}(\lambda(V))&>\max \lambda(V).
\end{align*}
Thus, $\max\lambda(V)\in\Theta(\operatorname{range}(\lambda(V)))$.
\end{Lemma}

\begin{proof}
Let $x\in V$ be the vertex carrying the smallest label $\min \lambda(V)$. As $x$ is not an isolate, there must be an edge incident to~$x$ that connects to a vertex $y$ such that
$\lambda(y)>\lambda(x)$. Hence, there must be a vertex $z$ in $V$ (possibly, an isolate) that carries a label $\lambda(z)=\lambda(x)+\lambda(y)>2\lambda(x)$.
Now, $\max \lambda(V)-\min\lambda(V)\geq \lambda(z)-\lambda(x)>\lambda(x)$.
Moreover, $2\cdot (\max \lambda(V)-\min\lambda(V))=(\max \lambda(V)-\min\lambda(V))+(\max \lambda(V)-\min\lambda(V))
>(\max \lambda(V)-\min\lambda(V))+\min\lambda(V)= \max\lambda(V)$.
\end{proof}
What is the main purpose of a graph database? Clearly, one has to access the graphs. A basic operation would be to answer the query if there is an edge between two vertices. 
Now, if $\max\lambda(V)$ of a sum graph is polynomial in the number $n=|V|$ of its vertices, we can answer this query in time $\Oh(\log(n))$, a property also discussed as~\emph{adjacency labelling scheme} by Peleg~\cite{Pel2000}.

Namely, assuming the polynomial bound on the size of the labels, we would need time $\Oh(\log(n))$
to add the two labels of the vertices, and we also need time $\Oh(\log(n))$
to search for the sum in the ordered list of numbers, using binary search, because there are only $\Oh(n^2)$ many numbers needed to describe a graph. If $\max\lambda(V)$ would be super-polynomial, then the additional time $\Oh(\log(\max\lambda(V)))$ would be quite expensive, which probably makes the idea of storing large graphs as sum graphs in databases unattractive. This motivates in particular also considering $\max\lambda(V)$ of the labelling $\lambda$ of a sum graph.

We discuss further graph storing schemes that may be thought of efficient in terms of their memory requirements in~\autoref{app:mwouslfsg}.



\paragraph{Lower Bounds.}
How many bits are really necessary to store graphs? We will discuss lower and upper bounds in the following, starting with a lower bound.


\begin{Lemma}\label{lem:lowerboundonn}
Let $G$ be an $n$-vertex graph. Then $\sigma_\textbf{store}^\text{max}(G)\in\Omega(n\log n)$, and $\sigma_\textbf{store}(G)\in\Omega(n\log n)$.
\end{Lemma}

\begin{proof} When it comes to storage costs, the most parsimonious labelling $\lambda:V\to\mathbb{N}$ obeys $\lambda(V)=[n]=\{1,2,\dots,n\}$ by injectivity. Now,
$$\sum_{v\in V}\lceil\log_2(\lambda(v))\rceil\geq \sum_{v\in V}\log_2(\lambda(v))=\sum_{i\in [n]}\log_2(i)=\log_2\left(\prod_{i\in [n]}i\right)=\log_2(n!)\,.$$
By Stirling's formula~\cite{dutka}, there are constants $c,d$ such that $$\log_2(n!)\geq \log_2(d\cdot (n/c)^n)=(n/c)\log_2(n)+\log_2(d)\in\Omega(n\log n)\,.$$
As $\sum_{v\in V}\lceil\log_2(\lambda(v))\rceil\leq |V|\max_{v\in V}\lceil\log_2(\lambda(v))\rceil$, both lower bound claims are true.
\end{proof}
This lemma shows that a sum labelling with $\Oh(n\log n)$ bits is storage-optimal, up to constants. This is one of the motivations underlying the discussions in the next section. Moreover, $\Omega(n\log n)$ is the space requirement for storing sparse graphs using traditional graph-storage methods. $\Omega(n\log n)$ bits are needed just to write the names of the vertices, as can be seen by a calculation similar to the proof of~\autoref{lem:lowerboundonn}.


\paragraph{Upper Bounds.} Here, we start our discussion on upper bounds for storing graphs with sum labellings. First, we briefly discuss the number of isolates in this respect. Based on some probabilistic arguments, it is known that the number of isolates is about the number of edges of the graph to be encoded~\cite{GouRod8891,NagMilSla2001} for nearly all graphs.

\begin{Remark}\label{rem-lowerbound}
As there are $2^{\Theta(n^2)}$ many graphs on $n$ vertices, we cannot hope for a sum labelling scheme that uses only $n^{2-\varepsilon}$ many isolates and only polynomial-size labels and hence a polynomial range, because we need at least $\Omega(n^2)$ many bits just to write down $n$-vertex graphs. As an aside, allowing for $n^2$ many isolates also means always allowing exclusive labellings.
\end{Remark}
Conversely, assuming we can sum-label each $n$-vertex, $m$-edge graph with polynomial-sized labels, then we can upper-bound $\sigma_\textbf{store}$ by $\Oh(m\log(n))$. By our discussions from~\autoref{lem:lowerboundonn} and~\autoref{rem-lowerbound}, we cannot hope for anything substantially better.
Can we reach this bound? Unfortunately, this seems to be an open question that we will answer to some extent below in our main result. In~\cite{KraMilNgu2001}, it was shown that each $n$-vertex graph without isolates can be represented by a sum labelling that uses numbers no larger than $4^n$. In other words, one would need at most $2n$ bits to represent each vertex of an $n$-vertex graph. This also shows that sum graphs have a~\emph{constrained 1-labelling scheme} as defined in~\cite{KanNaoRud92}.  Hitherto, it was unknown how to sum-label arbitrary graphs with polynomial-size labels. 
As our main result, we solve this problem affirmatively, with nice consequences for $d$-degenerate graphs.

\section{A Novel Algorithm for Sum Labelling}

We will now prove our main result (\autoref{thm-main}), thereby showing that sum labellings can be used to store graphs as efficiently as traditional methods can do. It is easy to see that the two major theorems shown in this section (\autoref{thm:labelsize-general},~\autoref{thm:labelsize-degenerate}) imply~\autoref{thm-main}.\footnote{For the sake of simplicity, in our proofs we assume that the given graphs have no isolates. It is easy to see that the same bounds hold (up to constant additive terms) even when the given graphs have isolates.}

\begin{Theorem}\label{thm:labelsize-general}
Every $n$-vertex, $m$-edge graph $G$ of minimum degree at least one can be made a sum graph $H$ by adding at most $m$ isolates to $G$, such that $H$ admits a sum labelling $\lambda$ satisfying
\begin{align}
    \lambda(v)&\leq 4\cdot n^3\quad&&\forall\,v\in V(G);\label{eq:n3ext1}\\
    \lambda(v)&\leq 8\cdot n^3\quad&&\forall\,v\in V(H).\label{eq:n3ext2}
\end{align}
Furthermore, the labelling is an exclusive sum labelling, computable in polynomial time by~\autoref{alg:algsumlab}.
\end{Theorem}

Our proof of~\autoref{thm:labelsize-general} is constructive and algorithmic in nature, described formally by~\autoref{alg:algsumlab}. The proof itself explains in words the working of this algorithm, its correctness, and provides upper bounds on the sizes of the vertex labels.

\paragraph{The Algorithm.} \autoref{alg:algsumlab} takes a non-empty graph $G$ as input, and outputs a sum graph $H$ and a labelling $\lambda$ of $H$ such that $H=G+\overline{K_c}$ (for some $c\leq m$) is a sum graph with sum labelling $\lambda$. \autoref{alg:algsumlab} uses~\autoref{alg:valsumgra} as a subroutine. \autoref{alg:valsumgra} takes a graph $H$ and a labelling~$\lambda$ of $H$ as input, and outputs $\mathsf{TRUE}$ if $\lambda$ is a sum labelling of $H$ and $\mathsf{FALSE}$ otherwise.\footnote{For technical reasons,~\autoref{alg:valsumgra} allows different vertices to have the same label as long as they are isolates. This is not allowed in sum labelling; in the end,~\autoref{alg:algsumlab} fixes this by eliminating isolates with duplicate labels.}

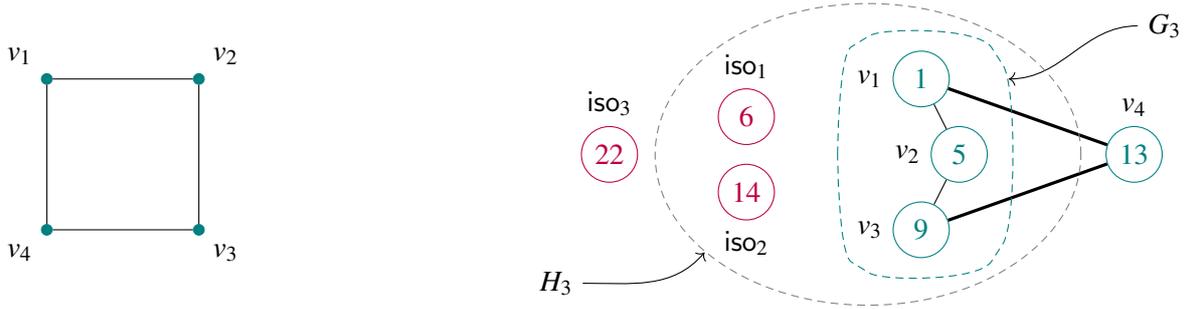
\begin{figure}
\begin{center}
\begin{tikzpicture}
    
    \def \vgap {0.6}
    \def \hgap {2.5}
    
    \vertex(v1) at (-8,4) [minimum size=4pt, teal, fill=teal, label=above left:$v_1$] {};
    \vertex(v2) at (-6,4) [minimum size=4pt, teal, fill=teal, label=above right:$v_2$] {};
    \vertex(v3) at (-6,2) [minimum size=4pt, teal, fill=teal, label=below right:$v_3$] {};
    \vertex(v4) at (-8,2) [minimum size=4pt, teal, fill=teal, label=below left:$v_4$] {};
    
    \draw (v1)--(v2)--(v3)--(v4)--(v1);
    
    \node[vertex, teal, label=left:$v_1$] (v1b) at (3.5,4) {$1$};
    \node[vertex, teal, label=left:$v_2$] (v2b) at (4,3) {$5$};
    \node[vertex, teal, label=left:$v_3$] (v3b) at (3.5,2) {$9$};
    
    \node[vertex, purple, label=above:$\iso_1$] (i1) at (1.2,3.5) {$6$};
    
    \node[vertex, purple, label=below:$\iso_2$] (i2) at (1.2,2.5) {$14$};
    
    \node[vertex, purple, label=above:$\iso_3$] (i3) at (-0.6,3) {$22$};
    
    \draw (v1b)--(v2b)--(v3b);
    
    \node[vertex, teal, label=above :$v_4$] (v4b) at (6.3,3) {$13$};
    
    \draw[very thick](v4b)--(v1b);
    \draw[very thick](v4b)--(v3b);
    
    \def \xoff {0.2};
    \def \yoff {-0.5};
    
    \draw [densely dashed, teal, rounded corners] plot [smooth] coordinates {
    (2.2+\xoff,3.5+\yoff)
    (2.5+\xoff,5+\yoff)
    (4.3+\xoff,4.8+\yoff)
    (4.3+\xoff,2.2+\yoff)
    (2.5+\xoff,2+\yoff)
    (2.2+\xoff,3.5+\yoff)};
    
    \node[anchor=south] at (-1.3,1) (h3) {$\Large H_3$};
	\draw (h3) edge[out=0,in=230,->] (0.65,1.7);
	
	\node[anchor=north] at (6.7,5) (g3) {$\Large G_3$};
	\draw (g3) edge[out=180,in=0,->] (4.65,4);
    
    \draw (2.8,3) [densely dashed, gray] circle [x radius=2.8cm, y radius=2cm];
    
    
    
\end{tikzpicture} \caption{(Left) The graph $G$ with the vertex ordering $\{v_1,v_2,v_3,v_4\}$ is provided as input to~\autoref{alg:algsumlab}. (Right) $G_3$ is the induced subgraph of $G$ on the vertex set $\{v_1,v_2,v_3\}$ and $H_3$ is its corresponding sum graph, constructed by the first three iterations of the algorithm, along with its labelling. At the fourth iteration, $v_4$ is added to $G_3$ to obtain $G_4$, and a new isolate $\iso_3$ is also added to $H_3$ to obtain $H_4$. Notice that the vertices of $G_4$ are labelled $1\,\operatorname{ mod }\,4$ and the isolates are labelled $2\,\operatorname{ mod }\,4$; this labelling scheme is described in the proof of~\autoref{thm:labelsize-general}. This specific example is also explained in more detail in~\autoref{ex:c4labelling}.}
\label{fig:optsol}
\end{center}
\end{figure}

\begin{proof}[Proof of~\autoref{thm:labelsize-general}] It is helpful to follow~\autoref{fig:optsol} while reading this proof. Notice that~\autoref{eq:n3ext1} implies~\autoref{eq:n3ext2}, as isolate labels are sums of labels of $V(G)$. So we will focus on showing~\autoref{eq:n3ext1} in this proof. Let the vertices of $G$ be $\{v_1,v_2,\ldots,v_n\}$. Let $G_i$ be the induced subgraph on the first~$i$ vertices of $G$, that is, $$V(G_i)=\{v_1,v_2,\ldots,v_i\}.$$ For each $G_i$ ($2\leq i\leq n$), we will show that there is a sum graph $H_i$ which can be obtained by adding  $r_i\leq \binom{i}{2}$ isolates to $G_i$ (since $G_i$ has at most $\binom{i}{2}$ edges), satisfying $\lambda(v)\leq 4\cdot i^3$ for each $v\in V(G_i)$. Moreover, all vertices of $G_i$ will carry labels that equal 1 modulo 4, and all isolates in $H_i$ will carry labels that equal 2 modulo 4. This modulo condition ensures that our labelling is exclusive. Our proof is by induction on $i$, yielding an algorithm explicitly described by~\autoref{alg:algsumlab}.

Although the statement of the theorem makes sense only from $n\geq 2$ onward to meet the minimum-degree requirement, it is convenient for our inductive proof to start with $i=1$:

\medskip\noindent\textbf{Base case ($i=1$):} We set $\lambda(v_1)=1$. Notice that $\lambda(v_1)=1^3$. Set $r_1=0$.

\medskip\noindent\textbf{Induction hypothesis:} There is a sum graph $H_i$ for $G_i$ such that $H_i$ has $r_i$ isolates (in other words, $H_i=G_i\cup\{\iso_1,\iso_2,\ldots,\iso_{r_i}\}$), where $r_i\leq\binom{i}{2}$, and $\lambda(v)\leq 4\cdot i^3$ for each $v\in V(G_i)$. Moreover, all vertices of $G_i$ carry labels that equal 1 modulo 4, and all isolates in $H_i$ carry labels that equal 2 modulo 4.

\medskip\noindent\textbf{Induction step:} We add the vertex $v_{i+1}$ to the graph $H_i$ and connect it to its neighbours in $G_i$. Suppose $v_{i+1}$ has $t_i$ neighbours $\{v_{j_1},v_{j_2},\ldots,v_{j_{t_i}}\}$ in $G_i$. Then  add $t_i$ isolates $\{\iso_{r+1},\iso_{r+2},\ldots,\iso_{r_i+t_i}\}$ to $H_i$, giving the graph  $H_{i+1}$. Thus, $$H_{i+1}=G_{i+1}\cup\{\iso_1,\iso_2,\ldots,\iso_{r_i+t_i}\}\,.$$
We define $r_{i+1}=r_i+t_i$. Next, we set the labels of the newly added vertices. If $\lambda$ is not a valid sum labelling for $H_{i+1}$, then we will change the $\lambda$-values of the newly added vertices. We will show that their $\lambda$-values need to be changed less than $i^3$ times until we reach a valid sum labelling for $H_{i+1}$. 
\begin{align}
    \lambda(v_{i+1})&=5;\label{eq:vit1}\\ 
    \lambda(\iso_{r_i+k})&=\lambda(v_{i+1})+\lambda(v_{j_k})\quad\forall\,k\in\{1,2,\ldots,t_i\}. \label{eq:vit2}
\end{align}

\begin{Claim} \label{cl:sumgraphifandonlyif}
$\lambda$ is a valid sum labelling of $H_{i+1}$ if and only if it has none of the following~\emph{violations}.
\begin{enumerate}
    \item[(i)] A violating pair: an ordered set of two vertices $(u,w)$ from $G_i$ such that $\lambda(u)=\lambda(w)$.
    \item[(ii)] A violating triple: an ordered set of three vertices $(u,w,y)$ such that $\lambda(u)<\lambda(w)<\lambda(y)$ and $\lambda(u)+\lambda(w)=\lambda(y)$ and $(u,w)\notin E(H_{i+1})$.
\end{enumerate}
\end{Claim}\label{cl:violations}

\begin{Remark} \label{rem:mod4arith}
Notice that it could happen that some of the `new' isolates in $H_{i+1}$ carry labels that are already labels of isolates from $H_i$. In that case, we implicitly delete the extra isolates ($t_i$ is decreased accordingly), automatically avoiding violating pairs among them. Also, the modulo 4 arithmetics prevent vertices of $G_i$ from pairing up with the isolates to form a violating pair.
\end{Remark}

\begin{proof}[Proof of~\autoref{cl:sumgraphifandonlyif}]
It is easy to see that if $H_{i+1}$ has any of the above violations, then $\lambda$ is not a valid sum labelling of $H_{i+1}$. Now we will prove the other direction: if $\lambda$ is not a valid sum labelling of $H_{i+1}$, then it either has a violating pair or a violating triple.

Notice that $H_{i+1}$ has $i+r_i+t_i+1=(i+1)+r_{i+1}$ many vertices, each with its corresponding $\lambda$-value. If two of the vertices have the same $\lambda$-value, then it is a type (i) violation, and we are done. So, we assume that all the $\lambda$-values are distinct. Given these $(i+1)+r_{i+1}$ distinct numbers, we construct their corresponding sum graph~$H'_{i+1}$ on $(i+1)+r_{i+1}$ vertices using the sum labelling property.

Both $H_{i+1}$ and $H'_{i+1}$ have the same set of vertices and the same labelling scheme $\lambda$. However, since $\lambda$ is a valid labelling scheme for $H'_{i+1}$ but not for $H_{i+1}$, they cannot have the same set of edges. Furthermore, $H_{i+1}$ is a subgraph of $H'_{i+1}$. This is because every edge $e=(u,w)$ of $H_{i+1}$ is either an edge that was also present in $H_i$ (in which case there is a vertex labelled $u+w$ in $H_{i+1}$ and $H'_{i+1}$, since $H_i$ is a sum graph by the induction hypothesis), or it is one of the $t_i$ new edges added (in which case one of the $t_i$ new isolates $\{\iso_{r_i+1},\iso_{r_i+2},\ldots,\iso_{r_{i+1}}\}$ is labelled $u+w$ by~\autoref{eq:vit2}).

Due to~\autoref{rem:mod4arith}, the only way for the edge sets of $H_{i+1}$ and $H'_{i+1}$ to differ is if there is an edge $e=(u,w)$ such that $e\in E(H'_{i+1})$ and $e\notin E(H_{i+1})$. This means there are three vertices $(u,w,y)$ in $H'_{i+1}$ (and so also in $H_{i+1}$) such that $\lambda(u)+\lambda(w)=\lambda(y)$, a type (ii) violation.
\end{proof}
Now, if $H_{i+1}$ is a sum graph with the labelling scheme derived from~\autoref{eq:vit1} and~\autoref{eq:vit2}, then we are done. Otherwise, we (slightly) modify these labels to obtain a new labelling, as follows.
\begin{align}
    \lambda(v_{i+1})&\leftarrow\lambda(v_{i+1})+4; \label{eq:revit1}\\
    \lambda(\iso_{r_i+k})&\leftarrow\lambda(\iso_{r_i+k})+4. \label{eq:revit2}
\end{align}
We again check if with these new labels, $H_{i+1}$ is a sum graph. If not, we increment these values by $4$ again. We keep doing this until $H_{i+1}$ becomes a sum graph. The crucial point to note is that each time we increment by $4$, at least one of the violations disappears, never to occur again.

To fully understand this last sentence, we need to refine our analysis of potential conflicts that might occur when running our algorithm. Namely, following up on the proof of the previous lemma, consider three vertices  $\{u,w,y\}$ in  $H_{i+1}$ such that (incorrectly) $\lambda(u)+\lambda(w)=\lambda(y)$ in the labelling $\lambda$ of $H_{i+1}$. First observe that not all vertices from $\{u,w,y\}$ can be isolates, as the isolates carry labels that are 2 modulo 4.

As we know that $\lambda$, restricted to the vertices of $H_i$, turns $H_i$ into a sum graph, not all of the vertices $\{u,w,y\}$ belong to $H_i$. If $y$ is one of the isolates of $H_i$, then its labelling will not change when updating $\lambda$ according to~\autoref{eq:revit2}. As one of the vertices $u,w$ does not belong to $H_i$, we have, w.l.o.g., $u\in V(H_i)$ and $w=v_{i+1}$, because if $w$ would be among the isolates, the sum of the labels of $u$ and $w$ would equal 0 modulo 4, but all isolates carry labels that are 2 modulo 4. This means that out of the three labels of $u,w,y$, exactly one will change according to~\autoref{eq:revit1} and as it will also be the only one that might increase in further modifications, a violation will never re-appear in the triple  $(u,w,y)$.

Assume now that $y$ is one of the new isolates, say, $y=\iso_{r_i+1}$. If exactly one of the two other vertices, say, $u$, already belongs to $H_i$, then the other one, $w$, must be $v_{i+1}$. As $\lambda(u)+\lambda(w)=\lambda(y)=\lambda(\iso_{r_i+1})$, we must have $u=v_{j_1}$, as we have no violating pairs. However, this means that the edge $(u,w)$ belongs to both $H_{i+1}$ and to $H_{i+1}'$, contradicting our assumption.
Therefore, if $y$ is one of the new isolates, then both $u$ and $w$ must belong to $H_i$. This means that the labellings of $u$ and of $w$ will never change by the re-labellings described in~\autoref{eq:revit1} and~\autoref{eq:revit2}, while the labelling of $y$ will only (further) increase, so that indeed a violation will never re-appear in the triple  $(u,w,y)$.

How often might we have to update a labelling when moving from $H_i$ to a valid sum graph $H_{i+1}$? Our previous analysis shows that the following are the only two scenarios that could possibly be encountered for a violating triple $(u,w,y)$:
\begin{enumerate}
    \item $y$ is an isolate of $H_i$ and exactly one of $\{u,w\}$ belongs to $V(G_i)$, while the other is $v_{i+1}$. There are at most $i\cdot r_i$ many cases when this might occur.
    \item $y$ is an isolate of $H_{i+1}$ and $\{u,w\}\subseteq V(H_i)$. There are at most $t_i\cdot\binom{i}{2}=t_i\cdot i(i-1)/2$ many cases when this might occur.
\end{enumerate}


Recall that $r_i$ isolates are contained in the sum graph $H_i$ and $t_{i}=r_{i+1}-r_i$ isolates are newly added to yield $H_{i+1}$.
Our analysis shows that after at most $s_i=i\cdot r_i+t_i\cdot i(i-1)/2$ many steps, a valid sum labelling of $H_{i+1}$ was found. By observing that $r_i$ cannot be larger than the number $\binom{i}{2}=i(i-1)/2$ of hypothetical edges in $H_i$, and $t_i$ is upper-bounded by the number $i$ of vertices in $H_i$, we can furthermore estimate:
$$s_i\leq i\cdot i(i-1)/2 + i\cdot i(i-1)/2= i^3-i^2\,.$$

By induction hypothesis, we know that for each of the $i$ vertices $v$ in $H_i$, we have $\lambda(v)\leq i^3$.  As $H_i$ contains only $i$ vertices that are labelled with number that are equal to 1 modulo 4, within at most $i^3-i^2$ increment steps, we will find a label for $v_{i+1}$ that is no larger than $4\cdot (i^3-i^2)+1\leq  (\sqrt[3]{4}\;(i+1))^3$,
basically using the pigeonhole principle. 
%
%
As all labels of isolates are sums of labels of vertices from $G_i$, their sizes are upper-bounded by $4i^3+4(i-1)^3<8\cdot i^3$.\end{proof}

This gives an upper bound of $(n+m)(\log(8 n^3))$ on the total number of bits required to store~$H$. Since every vertex in $G$ has degree at least one, we have $n\leq 2m$. Substituting, we get an upper bound of $3m(\log(8 n^3))\leq 3m(3\log n+3)=9m(\log n + 1)$, as required by~\autoref{thm-main}.

\begin{algorithm}
	\begin{algorithmic}[1]
        \State $V(H)\gets v_1$ \Comment{Initialising $H$}
        \State $\lambda(v_1)\gets 1$
        \State $E(H)\gets \emptyset$
        \State $c\gets 0$ \Comment{Counter for the number of isolates}
        \State $i\gets 2$ \Comment{The current vertex being processed}
        \While{$i\leq n$}
        \Comment{The ordering $V(G)=\{v_1,v_2,\ldots,v_n\}$ is part of the input}
            \State $V(H)\gets V(H)\cup\{v_i\}$
            \State $\lambda(v_i)\gets 5$\Comment{See~\autoref{eq:vit1}}
            \State $\cbef\gets c$
            \For{each $j$ such that $1\leq j\leq i-1$ and $v_iv_j\in E(G)$}
                \State $V(H)\gets V(H)\cup\{\iso_{c+1}\}$
                \State $E(H)\gets E(H)\cup\{v_iv_j\}$
                \State $\lambda(\iso_{c+1})\gets \lambda(v_i)+\lambda(v_j)$\Comment{See~\autoref{eq:vit2}}
                \State $c\gets c+1$
            \EndFor
            \State $\caft\gets c$
            \While{$\textsc{CheckValidSumGraph}(H,\lambda)=\mathsf{FALSE}$}
                \State $\lambda(v_i)\gets\lambda(v_i)+4$\Comment{See~\autoref{eq:revit1}}
                \For{each $\ell$ such that $1+\cbef\leq\ell\leq\caft$}
                    \State $\lambda(\iso_\ell)\gets\lambda(\iso_\ell)+4$\Comment{See~\autoref{eq:revit2}}
                \EndFor
            \EndWhile
        \EndWhile
        \For{each $(\iso_i,\iso_j)\in V(H)\times V(H)$ such that $i<j$}
        \If{$\lambda(\iso_i)=\lambda(\iso_j)$}
            \State $V(H)\gets V(H)\setminus\{\iso_j\}$ \Comment{Remove isolates with duplicate labels}
        \EndIf
        \EndFor
        \State \Return $(H,\lambda)$
    \end{algorithmic}\caption{$\textsc{SumLabel}(G)$}\label{alg:algsumlab}
\end{algorithm}

\begin{algorithm}
	\begin{algorithmic}[1]
	    \State $f\gets\mathsf{TRUE}$ \Comment{$f=\mathsf{TRUE}$ $\Leftrightarrow$ $H$ is a sum graph with sum labelling $\lambda$}
        \For{each $(v_1,v_2)\in V(H)\times V(H)$ such that $v_1\neq v_2$}
            \State $s\gets\mathsf{FALSE}$ \Comment{$s=\mathsf{TRUE}$ $\Leftrightarrow$ $v_1v_2$ is an edge as per the sum labelling $\lambda$}
            \If{$(\deg(v_1)\neq 0 \vee \deg(v_2)\neq 0) \wedge (\lambda(v_1)=\lambda(v_2))$}
                \State $f\gets\mathsf{FALSE}$ \Comment{Two vertices cannot have the same label, unless they are both isolates}
            \EndIf
            \For{each $v_3\in V(H)$}
                \If{$\lambda(v_3) = \lambda(v_1)+\lambda(v_2)$}
                    \State $s\gets\mathsf{TRUE}$ \Comment{The sum labelling says that $v_1v_2$ is an edge}
                \EndIf
            \EndFor
            \If{$(v_1v_2\in E(H)\wedge s=\mathsf{FALSE})\vee(v_1v_2\notin E(H)\wedge s=\mathsf{TRUE})$}
                \State $f\gets\mathsf{FALSE}$ \Comment{The sum labelling $\lambda$ does not concur with the graph $H$}
            \EndIf
        \EndFor
        \State \Return $f$
    \end{algorithmic}\caption{$\textsc{CheckValidSumGraph}(H,\lambda)$}\label{alg:valsumgra}
\end{algorithm}

\paragraph{Some Concrete Examples.} We now look at how~\autoref{alg:algsumlab} performs on some small graphs.

\begin{Example}\label{ex:k4labelling}
Let $\{v_1,v_2,v_3,v_4\}$ be the vertices of $K_4$. We label $\lambda(v_1)=1$, $\lambda(v_2)=5$ and introduce the isolate $\iso_1$ with $\lambda(\iso_1)=6$. Then, we label $\lambda(v_3)=9$, and introduce the isolates $\iso_2$, $\iso_3$ with $\lambda(\iso_2)=10$, $\lambda(\iso_3)=14$. Next, we label $\lambda(v_4)=13$. In principle, we would now introduce three isolates with labels $13+1$, $13+5$, $13+9$. But, as the label 14 is already present for $\iso_3$, we need only two new isolates with labels $18$, $22$. In this way, our labelling scheme even finds the optimal sum labelling for $K_n$ in general. Incidentally, this labelling scheme also gives a labelling of minimum spum (range).
\end{Example}
\begin{Example} \label{ex:c4labelling}
For labelling a $C_4$ whose vertices are $(v_1,v_2,v_3,v_4)$ in cyclic order (see~\autoref{fig:optsol}), the first two steps are the same (i.e., $\lambda(v_1)=1,\lambda(v_2)=5,\lambda(\iso_1)=6$) as in \autoref{ex:k4labelling}, but after setting $\lambda(v_3)=9$, the second isolate $\iso_2$ is labelled $\lambda(\iso_2)=14$. This describes the edges $v_1v_2$ and $v_2v_3$. Now $v_4$ enters the scene, with edges to $v_1$ and to $v_3$. When using $\lambda(v_4)=13$, the edge $v_1v_4$ is already properly labelled by $\iso_2$. With a third isolate $\iso_3$ labelled $\lambda(\iso_3)=22$, we again find an optimal sum labelling, since we know that $\sigma(C_4)=3$.

However, this was a bit lucky: if the cyclic order was $(v_1,v_2,v_4,v_3)$, then we would have $\lambda(v_3)=9$ and $\lambda(\iso_2)=10$. Now, $\lambda(v_4)=13$ would lead to isolates labelled $\lambda(\iso_3)=18$ and $\lambda(\iso_4)=22$, so we would actually need four isolates in this case.
\end{Example}

As we always start with setting the label of the first vertex to $1$,  the obtained labelling uses the number $1$ as a label.
Notice that this is related to the
(to the best of our knowledge, still open) question whether every graph $G$ (without isolates) can be embedded into a sum graph $H$ with $\sigma(G)$ many isolates such that there is a sum labelling $\lambda$ of $H$ with $\lambda(v)=1$ for some vertex $v\in V(H)$, see
~\cite{MilRyaSmi98,Kone}.

\paragraph{Modifications of our Algorithm.} Notice that we are creating new isolates only when necessary. This has the nice consequence that we can use the same isolate for various edges. Due to this, our algorithm recovers the optimal sum labelling of the complete graph $K_n$, for example.

However, there are circumstances when this kind of optimization is not really wanted. For instance, when we store graphs that behave more dynamically, we might want to have the possibility to quickly  delete edges. In that case, it is beneficial to use exactly $m$ distinct edge labels (i.e., isolates) to help with these updates, as then, no further changes or re-computations of vertex labels are necessary, as only the respective isolates have to be deleted. Similarly, vertex deletions can be incorporated efficiently.

We can modify our algorithm to ensure that (new) vertex labels are changed (according to~\autoref{eq:revit1} and~\autoref{eq:revit2}) until this uniqueness condition concerning edge labels is satisfied. By using the same pigeonhole argument, the overall argument of the algorithm is not changed, so that we can even meet that upper bounds on label sizes promised in~\autoref{thm:labelsize-general}
and~\autoref{thm:labelsize-degenerate} for this modification. 

As our labelling algorithm can be thought of building up the graph vertex-by-vertex, also adding vertices to an existing, labelled graph is not that difficult, because we can simply run our algorithm one step further, this way processing the new vertex (and its incident edges).

It is not that clear if we can further modify our algorithm to also cope with edge additions, as this might require re-labelling the vertices.
All these discussions respond to the question if and how sum labellings can be used for storing and accessing possibly dynamically evolving graphs.

A further natural modification of our algorithm would be a randomized variation thereof.\footnote{This idea was given by Jaikumar Radhakrishanan during a (virtual) talk on this work at TIFR in October, 2021.} At first thought, one might think that by selecting a random number in a certain interval, there is a good chance to pick a number that produces the required edges and avoids any unwanted ones. However, our thoughts in this direction revealed that this interval should be a range of numbers in $\{1,2,\ldots,n^6\}$ or a similar polynomial upper bound. This is obviously worse than what we could achieve with our deterministic algorithm. Yet, further improvements of a randomized algorithm might be possible and could then lead to some ideas of storing graphs that are better suited for update operations on graphs.

\section{Labelling Sparse Graphs}
We will now look into specific classes of sparse graphs. We consider graph degeneracy as our primary measure of sparseness. Notice that sparse graphs are often considered as modelling real-world networks more faithfully than general graphs that could be arbitrarily dense. In fact, as we will see, this restriction does give us some advantage when storing graphs with sum-labelling schemes.

\begin{Definition}
A graph is called $d$-degenerate if every subgraph of the graph has a vertex of degree at most $d$. The degeneracy of a graph is the minimum $d$ for which it is $d$-degenerate.
\end{Definition}

It is easy to see that the vertex set of a $d$-degenerate graph $G=(V,E)$ can be ordered as $V=\{v_1,v_2,\ldots,v_n\}$ in polynomial time such that the vertex $v_i$ has degree at most~$d$ in the graph $G_i$ induced by the vertices $V_i=\{v_1,v_2,\ldots,v_i\}$. We call such an ordering a~\emph{$d$-degenerate vertex ordering}. We will use this concept in the proof of the following theorem. 

\begin{Theorem}\label{thm:labelsize-degenerate}
Every $d$-degenerate, $n$-vertex, $m$-edge graph $G$ of minimum degree at least one can be made a sum graph $H$ by adding at most $m$ isolates to $G$, such that $H$ admits a sum labelling $\lambda$ satisfying
\begin{align}
    \lambda(v)&\leq 6d \cdot n^2\quad&&\forall\,v\in V(G);\label{eq:n3ext1degen}\\
    \lambda(v)&\leq 12d\cdot n^2\quad&&\forall\,v\in V(H).\label{eq:n3ext2degen}
\end{align}
This sum labelling is an exclusive labelling, computable in polynomial time.
\end{Theorem}

\begin{proof}
We will only point to the changes needed to make the analysis of~\autoref{thm:labelsize-general} work in this special case.
Recall that~\autoref{thm:labelsize-general} was proved by induction 
on an~\emph{arbitrary} ordering of its vertex set~$V$. However, in this proof, since $G$ is $d$-degenerate, we pick a $d$-degenerate vertex ordering $V=\{v_1,v_2,\dots,v_n\}$ of $G$. Recall that $G_i=G[\{v_1,v_2,\dots,v_i\}]$. For $G_i$, a sum graph $H_i$ was constructed by adding $r_i$ isolates. We add the following assertions that we are going to prove inductively about~$H_i$:
\begin{itemize}
    \item $H_i$ contains  $r_i\leq d\cdot (i-1)$ many isolates that are not vertices of $G_i$.
    \item For labelling vertices of $G_i$, labels no larger than $6d\cdot i^2$ are used.
    \item For labelling isolates of $H_i$, labels no larger than $12d\cdot i^2$ are used.
\end{itemize}
Moreover, the vertex $v_{i+1}$ added to $G_i$ in order to obtain $G_{i+1}$ has $t_i\leq d$ many neighbours in $V(G_i)$, as guaranteed by a $d$-degenerate vertex ordering. Now, in the analysis of the induction step, the main point was to discuss two cases of a violating triple $(u,w,y)$.
\begin{itemize}
    \item $y$ is an isolate of $H_i$ and exactly one of $\{u,w\}$ belongs to $V(G_i)$, while the other is $v_{i+1}$. There are at most $i\cdot r_i\leq d\cdot i\cdot (i-1)$ many cases when this might occur.
    \item $y$ is an isolate of $H_{i+1}$ and $\{u,w\}\subseteq V(G_i)$. There are at most $t_i\cdot i(i-1)/2\leq d\cdot i\cdot (i-1)/2$ many cases when this might occur.
\end{itemize}
This proves that after at most $s_i=\frac32 d\cdot i\cdot (i-1)$ many increment steps, $v_{i+1}$ will have a label no larger than $6d\cdot i^2$. This also proves the claimed bound on the label size for the isolates.
\end{proof}

This gives an upper bound of $(n+m)(\log(12 dn^2))$ on the total number of bits required to store $H$. Since every vertex in the graph $G$ has degree at least one, we have $n\leq 2m$. Substituting, we get an upper bound of $3m(\log(12 dn^2))\leq 3m(2\log n+\log 12d)$, as required by~\autoref{thm-main}.

\paragraph{Labelling Planar Graphs.} Since planar graphs are 5-degenerate~\cite{lick}, our sum labelling needs labels with $2\log_2(n)+\Oh(1)$ bits for storing planar graphs (by taking logarithms in~\autoref{eq:n3ext2degen}), improving on previous published bounds for implicit representations of planar graphs~\cite{Sch89,Sch90,BonGHPS2006,GavLab2007,KanNaoRud92,KeeWes95,MunRam2001}, except the very last proposal~\cite{BonGavPhi2020} (see also~\cite{DujEGJMM2020}).

In adjacency labelling, the labels of two vertices alone are enough to decide whether the vertices are adjacent or not; for sum labelling, one needs to additionally check the labels of all the other vertices. Thus, sum labelling is not an adjacency labelling. However, our approach generalises to graphs of arbitrary fixed degeneracy, which is unclear for other approaches from the literature on adjacency labelling schemes.

In a recent breakthrough~\cite{DujEGJMM2020}, it was shown that for every $n$, there is a ``universal graph'' $U_n$ on $n^{1+o(1)}$ vertices such that every $n$-vertex planar graph is an induced subgraph of~$U_n$. Analogously,~\autoref{thm:labelsize-degenerate} implies that every $n$-vertex planar graph can be represented by a subset of $[60n^2]$ (as planar graphs are $5$-degenerate and our upper bound is $12dn^2$). Is it possible to arrive at sum labelling representations for planar graphs that only need numbers from $[c\cdot n^{1+o(1)}]$ instead, for some constant $c$? This open question is a bridge to the final section, where we also discuss several lines of future research in this area.

\section{Discussion}


It is an interesting question how bad the labelling produced by our algorithm could get if it comes to determining the exclusive sum number of a graph.
To give another example, when labelling the complete bipartite graph $K_{|P|,|Q|}$, with its vertex set~$V$ split into two independent sets $P$, $Q$, the ordering that first lists $P$ and then $Q$ will actually produce the optimal exclusive sum labelling as suggested in~\cite{MilPRSST2005,Rya2009}.
Also by presenting the vertices of $P$ and $Q$ alternatingly to our algorithm, one can produce a labelling that realizes the exclusive sum number $|P|+|Q|-1$ of $K_{|P|,|Q|}$, but then the range is nearly twice as large. 

This brings us to the following interesting question: is there always a vertex ordering such that our algorithm yields an optimal exclusive sum labelling?

\begin{Proposition}\label{prop:good+bad-labellings}
There exists a family of graphs $(G_n)$ such that, if our algorithm is presented with a certain ordering of $V(G_n)$, where $|V(G_n)|=n\geq 3$, then it will produce a labelling $\lambda_n$ matching $\epsilon(G_n)$, but if presented with a different ordering, it will yield a labelling $\lambda_n'$ requiring $|E(G_n)|$ many isolates. The ratio between the number of isolates produced by $\lambda'_n$ and $\epsilon(G_n)$ grows beyond any limit.
\end{Proposition}

\begin{proof}The mentioned family of graphs is the family of paths. The exclusive sum number of paths equals two. Let us check this first with a small example: let us discuss $1-2-3-4-5$ as a $P_5$. However, given the ordering $1,2,3,4,5$ of the vertices, our algorithm would produce the labelling $\lambda(1)=1$, $\lambda(2)=5$, $\lambda(3)=9$, $\lambda(4)=17$, $\lambda(5)=29$, with the isolates labelled $6,14,26,46$. In general, presenting the vertices in such a sequence would require $n-1$ isolates for an $n$-vertex path, which is as bad as it could be in terms of the number of isolates. Yet, the ordering $1,3,5,4,2$ gives $\lambda(1)=1$, $\lambda(3)=5$, $\lambda(5)=9$, $\lambda(4)=13$, $\lambda(2)=17$, with only two isolates (which is optimal), $18$ and $22$. This is also true in general: if the vertices $1-2-\cdots-n$ of a $P_n$ are presented as $1,3,\dots,n,n-1,n-3,\dots,2$ (if $n$ is odd) or as $1,3,\dots,n-1,n,n-2,\dots,2$ (if $n$ is even), then an optimal exclusive sum labelling is achieved, with the isolates labelled $4n-2$ and $4n+2$ (if $n$ is odd) or $4n+2$ and $4n+6$ (if $n$ is even).
\end{proof}
As shown in this proof, the family of paths on $n$ vertices gives such a graph family. The labelling that is optimal  with respect to the exclusive sum number is different from the one proposed in~\cite{MilPRSST2005,Rya2009}.

Moreover, the following computational complexity questions are of interest, in particular, if one wants to apply sum labellings for storing real-world graphs. Are there polynomial-time algorithms for (any of) the following questions, given a graph $G$ without isolates?
\begin{itemize}
    \item Determine the sum number $\sigma(G)$ and find a corresponding sum labelling.
    \item Determine the exclusive sum number $\epsilon(G)$ and find a corresponding exclusive sum labelling.
    \item Find a sum labelling minimizing the range of the labels.
    \item Find a sum labelling minimizing the storage needs $\sigma_\textbf{store}^\text{max}(G)$ or $\sigma_\textbf{store}(G)$.
\end{itemize}

In particular, if a question of the suggested form would be $\NP$-hard, it would be interesting to know if there are good heuristics that order the vertices of a graph in a way that our algorithm produces a provable approximation to the best graph parameter  value. As the proof of~\autoref{prop:good+bad-labellings} shows, for instance the strategy behind the proof of~\autoref{thm:labelsize-degenerate} would actually produce a worst-case labelling in a sense, i.e., even labellings that have some good properties can be really bad with respect to another criterion. If it comes to giving an $\NP$-hardness proof for any of these questions, one of the difficulties is that the graph parameters related to sum labelling have a non-local flavour in the sense that local modifications of a graph could have tremendous effect on the graph parameters. It seems important to further study different typical graph operations with respect to these parameters. Here, more results like~\autoref{lem:disjointiterativeunion} are needed~\cite{KorPelRod2006}.

Among the hundreds of different graph labellings presented in~\cite{Gal2020}, the following are closest to sum labellings and could lead to considerations similar to the ones of this paper.
\begin{itemize}
    \item Integral sum labellings~\cite{Har94}, where also negative numbers are allowed to be used for labelling;
    \item Modulo (mod) sum labellings~\cite{harary,SutMil99,SutMRS99,Sut2000}, where addition modulo $k$ is used as operation on natural numbers; \item Product labellings~\cite{BerHJKWH92}, where the product operation on natural numbers is used instead of the summation.
\end{itemize}
These labellings can also be used to store graphs. Hence, questions similar to the ones raised and partially answered in this paper for sum labellings could be also considered for other graph labellings. Notice that although sum and product graphs coincide~\cite{BerHJKWH92}, the sizes of the labels are quite different, and therefore different labellings might have their own pros and cons if it comes to storing graphs. Seen from  a computer science perspective, it would even make sense to look at further labelling schemes not (yet) considered in the graph theory literature, for instance, mapping vertices to bit-vectors and then storing edges by means of bit-vectors obtained by, say, a bitwise OR-operation or AND-operation, because such operations can be implemented quite efficiently, similar to addition, and better than, say, multiplication, which is likely to be the least interesting number operation in our context anyways.

All these questions could open up quite new and challenging lines of research, possibly 
also further bridging to adjacency labellings~\cite{KanNaoRud92,Pel2000}.

Finally, recall that the sum labelling for trees proposed in~\cite{ellingham} introduces labels of exponential size. This means that, although only one isolate is added (proving that trees have sum number one), at worst $\Omega(n^2)$ many bits might be needed to encode trees in this way, while our approach needs only $\Oh(n\log(n))$ bits to store trees, using at worst $n-1$ many isolates, as shown in the proof of \autoref{prop:good+bad-labellings} for the case of paths. It is an open question if there is a sum labelling of any $n$-vertex tree~$T$ that uses $\Oh(n\log(n))$ bits and still certifies that $\sigma(T)=1$. A similar question can be asked concerning exclusive labellings, aiming at matching $\epsilon(T)$. However, to the best of our knowledge, no general formula is known for $\epsilon(T)$. The most interesting fact in this direction was proved in~\cite{Rya2009} for caterpillar graphs: here, the exclusive sum number matches the maximum degree. Also, the given labelling only uses labels of polynomial size.

One of our motivations to return towards sum labellings of graphs was the possibility to store graphs in a database.
We already discussed above that the question if an edge is present or not can be efficiently answered with sum label representations.
We have discussed several operations (like accessing adjacency information, and adding or removing vertices or edges) above in the context of our algorithm. In particular, an ever-expanding database that is gradually built up can be efficiently implemented and then accessed using our sum-labelling scheme. 

\paragraph{Acknowledgements.} The authors are grateful to the organisers of \textsc{Graphmasters 2020}~\cite{gkasieniec2020} for providing the virtual environment that initiated this research. A part of this work was done when the second author was a postdoctoral researcher at Technion, Israel. This project has received funding from the European Union’s Horizon 2020 research and innovation programme under grant agreement No. 682203-ERC-[Inf-Speed-Tradeoff]. An extended abstract of this paper appeared in~\cite{FerGaj2021}.

\bibliographystyle{alpha}

\bibliography{references}

\appendix












\section{More Ways of Using Sum Labellings for Storing Graphs} \label{app:mwouslfsg}


In~\cite{KraMilNgu2001}, the authors started out with optimal sum labellings and showed that a sum labelling of a sum graph $H=G+\overline{K_{\sigma(G)}}$ is possible with label sizes at most $4^{|V(G)|+\sigma(G)}$. 
This is not what we do in our theorems. Hence, there might be a trade-off between label sizes and number of isolates. We further on this in the next remark. But we must make it clear that our theorems (unfortunately) do not solve Problem~1 as formulated in~\cite{KraMilNgu2001} where the question was asked if one can prove a bound of $o(4^n)$ for the size of labels needed to realize $\sigma(G)$ for every $n$-vertex sum graph $G$.

In the following, we discuss an alternative approach for sum labelling arbitrary graphs. As we will see, this might also lead to graph representations that need $\Omega(m\log(n))$ many bits. However, it is not clear if fast graph queries are possible with this representation.

\begin{Theorem} \label{thm:incimatr}
Every graph on $n$ vertices without isolates can be embedded into a sum graph with at most $\frac{1}{2}n^2$ many isolates, so that each of the labels (also for the isolates) does not need more than 
$n+2$ many bits. Hence, the numbers involved in labelling $n$-vertex graphs grow with $\Oh(2^n)$. In addition, the resulting labelling is exclusive.
\end{Theorem}

\begin{proof}
Let $G$ be an $n$-vertex, $m$-edge graph. Consider its $n\times m$ incidence matrix $I_G$. Each column gives a bit-vector for an edge.
Now, first add two rows to this matrix $I_G$ on top,  an all-ones row, followed by an all-zeroes row; this gives the matrix $I_G'$, with $(n+2)$ rows and $m$ columns. Then, consider the $n\times n$ identity matrix $I$. Append two rows to on top: an all-zeros row, followed by an all-ones row, and call this matrix $I'$, giving an $(n+2)\times n$ matrix. Now, concatenate these matrices: first put the columns of $I'$, followed by the columns of $I_G'$, to get an $(n+2)\times (n+m)$ matrix $S_G$. The columns of $S_G$ are treated as binary numbers that label the vertices of $G$ (in the part $I'$) and that store the labels of the edges of $G$ (as we aim at an exclusive labelling).
Notice the importance of the leading two bits (the two rows that are added on top): when we add two bit-vectors denoting edges, we will create a number that is not in the list of vertex labels (as twice the highest bit is set). Also, when adding a bit-vector denoting an edge and a vertex, we create a bit-vector that starts with $11$, which is not found among the vertex labels. Finally, by construction, all labels needed for the isolates designating edges in~$G$ are present, as we started with the incidence matrix of $G$.
\end{proof}

When it comes to storing graphs, the $n+m$ many bit-vectors in the proof of~\autoref{thm:incimatr} can be stored more efficiently than using $(n+2)(n+m)$ many bits, because the column vectors contain at most three 1-bits each. Hence, one would need \begin{align}2\lceil \log_2(n)\rceil +1+2\lceil \log_2(m)\rceil +1 + n\lceil\log_2(n)\rceil+2m\lceil \log_2(n)\rceil\nonumber\\=(n+2m+2)\lceil \log_2(n)\rceil+2\lceil \log_2(m)\rceil+2\in\Oh(m\log(n))\label{eq:compressed} \end{align}
many bits to store first the number of vertices (in the format $1^{\lceil \log_2(n)\rceil}0(n)_2$, where $(x)_2$ refers to a binary string for the number $x$), then the number of edges (in the same format) and finally $n$ pointers signalling the 1-bit of the vertex labels and $2m$ pointers signalling the two 1-bits of the vertex labels.
This is much better than the $\Oh(mn)$, i.e., 
$$(n+m)(n+2)=n^2+nm+2n+2m\,,$$
bound for the number of bits in the original form of the previous theorem.
However, it is not that clear if one can query graph edges as efficiently (as discussed earlier for sum labelling formats).

One might argue that the term $m$ looks bad, but at least on average $m$ is about the sum number of a graph, see~\cite{NagMilSla2001}. So, only for special cases (as the $K_n$ shows), $\sigma(G)$ is way smaller, and then the result~\cite{KraMilNgu2001} that uses at most $$(n+\sigma(G))\cdot \log_2(4^{n+\sigma(G)})=2(n+\sigma(G))^2$$ many bits is superior to ours. More precisely, for storing $K_n$, the algebraic method of Kratochv\'il, Miller and Nguyen uses at most  $18n^2$ many bits, while our original method would need $(n+n^2/2)(n+2)$ many bits, while our improved method needs
about $(n+n^2+2)\log_2(n)+4n$ many bits according to~\autoref{eq:compressed}.

In fact, this calculation shows that (up to logarithmic factors), that our compressed encoding is at least as good as the encoding offered by~\cite{KraMilNgu2001}.
Compared to traditional ways of storing graphs, observe that with an adjacency matrix, one would need $n^2$ many bits, while with an adjacency list, one needs (again)  $\Oh(m\log(n))$ many bits, which is comparable with~\autoref{eq:compressed}.

\paragraph{A Novel Variant of Sum Labelling.} We already discussed exclusive sum labellings as a stricter variant of sum labelling. We are now proposing a relaxation of the notion.
 Namely, another way of using the concept of sum labelling for storing graphs is when the newly added vertices are not constrained to be isolates, leading to the notion of~\emph{supersum labelling}. That is, given a graph $G$, let $\varsigma(G)$ be the minimum number of (not necessarily isolated) vertices that need to be added to $G$ to make it a sum graph. In other words, we are looking for the smallest sum graph~$H$ that is a supergraph of a given graph~$G$ and $G$ is an induced graph of $H$.
Clearly, $\varsigma(G)\leq \sigma(G)$. In fact, there are examples where $\varsigma(G)<\sigma(G)$. For instance, we know that for the 4-cycle, $\sigma(C_4)=3$. We can show that $\varsigma(C_4)\leq 2$ by labelling its vertices $(1,2,3,5)$ in cyclic order, and the two additional vertices $6$ and $8$. (This is not allowed in sum labelling since $(2,6,8)$ would be a violating triple.) We are going to explore this new graph parameter in a subsequent paper.

As a final note, one could also define a supersum variation of exclusive sum labelling, leading to the graph parameter $\varsigma_\epsilon$, meaning that no vertex in the encoded graph $G$ is a working vertex of the labelling. 
However, this would not be an interesting field of future research, as our next (and final) theorem proves.

\begin{Theorem}
Let $G$ be a graph without isolates. Then, $\epsilon(G)=\varsigma_\epsilon(G)$.
\end{Theorem}

\begin{proof}
By definition, $\epsilon(G)\geq \varsigma_\epsilon(G)$. Let $\lambda$ be an exclusive labelling of $G$ that certifies $\varsigma_\epsilon(G)$. Now, consider the labelling $\lambda'$ that labels $v\in V(G)$ by $4\cdot \lambda(v)+1$.
Let $H$ be the sum graph realizing $G$ and $\lambda$, and also $G$ and $\lambda'$. For the vertices $v$ in $V(H)=V(H')$ that are not in $V(G)$, we have $\lambda'(v)=4\cdot \lambda(v)+2$. By modulo-4 arithmetics, it is rather obvious that $\lambda'$ is an exclusive sum labelling such that there are no edges in the sum graph $H'$ (realizing $G$ and $\lambda'$) between vertices not from $G$. Hence, $\epsilon(G)\leq \varsigma_\epsilon(G)$.
\end{proof}
Hence, at best  two bits per vertex and edge could be saved with a $\varsigma_\epsilon$-sum labelling, compared to the classical exclusive labelling. Also, one would need an algorithm different from the one proposed in this paper to exploit the fact that supersum labellings could be more parsimonious than sum labellings.

\end{document}